%% file: main.tex
\documentclass[lettersize,journal]{IEEEtran}
\IEEEoverridecommandlockouts
\usepackage{amsmath,amsfonts}
\usepackage{algorithmic}
\usepackage{algorithm}
\usepackage{subcaption}
\IEEEoverridecommandlockouts

\usepackage{bbm}
\usepackage{graphicx}

\usepackage{xcolor}
\usepackage{float}   
\usepackage[normalem]{ulem}
\usepackage{dsfont}
\usepackage{textcomp}
\usepackage{stfloats}
\usepackage{url}
\usepackage{verbatim}
\usepackage{algorithmic}
\usepackage{array}
\usepackage{booktabs}
\usepackage{pifont}
\usepackage{comment}
\usepackage{algorithm}
\usepackage{algorithmic}
\usepackage[acronym]{glossaries}
\input{acronyms}
\usepackage{amsmath,amsfonts, amssymb}
\usepackage{bm}
\usepackage{graphicx}
\usepackage{cite}

\usepackage[font=small]{caption}
\usepackage[font=footnotesize]{subcaption}
\usepackage{tikz}
\usetikzlibrary{plotmarks,patterns,decorations.pathreplacing,backgrounds,calc,arrows,arrows.meta,spy,matrix}
\usepackage{tikzscale}

\usepackage{pgfplots}
\usepgfplotslibrary{colorbrewer, groupplots, patchplots}
\pgfplotsset{
    compat=newest,
    legend style={font=\footnotesize, fill opacity=0.7,  draw opacity=1, text opacity=1, draw=white!15!black, legend cell align=left, align=left}, 
    width=0.8\columnwidth, 
    scale only axis,
    height=4cm,
    yminorticks=false,
    xminorticks=false,
    label style={font=\small},
    title style={font=\small},
    tick align=outside,
    tick pos=left,
    tick style={color=black},
    tick label style={font=\footnotesize},
    grid style={line width=.1pt, draw=gray!20},
    major grid style={line width=.1pt,draw=gray!20},
    plot coordinates/math parser=false 
}
\newlength\figureheight
\newlength\figurewidth

\usepackage{multirow}
\usepackage{tablefootnote}
\usepackage{booktabs}
\usepackage{tabularx}

\def\BibTeX{{\rm B\kern-.05em{\sc i\kern-.025em b}\kern-.08em T\kern-.1667em\lower.7ex\hbox{E}\kern-.125emX}}

\newcommand{\probP}{\text{I\kern-0.15em P}}

\definecolor{amaranth}{rgb}{0.9, 0.17, 0.31}
\newcommand{\BLUE}[1]{\textcolor{black}{#1}}

\begin{document}

\title{Energy-Efficient Timely Relevant IoT Data Retrieval with Wireless Memory Approximation}
\author{Junya Shiraishi,~\IEEEmembership{Member,~IEEE},
Zihang Song,~\IEEEmembership{Member,~IEEE}, Morten Andersen,~\IEEEmembership{Graduate Student Member,~IEEE}, and Petar Popovski,~\IEEEmembership{Fellow,~IEEE}\\
\thanks{
The work was supported by Horizon Europe Marie Sk{\l}odowska-Curie Action Postdoctoral Fellowship (MSCA-PF), with grant No.~101151067 and grant No. 101206861, and Villum Investigator Grant WATER, nr. 37793. (Corresponding author: J. Shiraishi).} 
\thanks{Z. Song, M. Andersen, and P. Popovski are with the Department of Electronic Systems, Aalborg University, 9220 Aalborg, Denmark, where J. Shiraishi was affiliated (e-mail: \{zsong, mortena, petarp\}@es.aau.dk); J. Shiraishi is now with National Institute of Information and Communications Technology, Yokosuka, Kanagawa, 239-0847 Japan (e-mail: jush@nict.go.jp). }}

\maketitle
\begin{abstract}
Collecting relevant data from Internet of Things devices under freshness and energy constraints is a key challenge in 6G edge networks. A lightweight neural network on the device filters irrelevant observations, but storing it in dynamic random access memory consumes refresh energy during standby. This letter proposes a pull-based framework where an edge server remotely activates and approximates memory and jointly selects the activation timing, refresh period, model size, and transmission threshold. We derive the expected freshness and energy over erasure channels in closed form and show that the scheme consumes less energy than an always-on baseline under the same freshness constraint.
\end{abstract}
  \vspace{-4mm}
\section{Introduction} 
\Gls{iot} networks collect multimodal sensory data to support emerging applications such as wildlife monitoring~\cite{kalor2025data} and the remote control of mobile robots~\cite{bezerra2026ai}. Because such data has complex patterns, a lightweight \gls{nn} on the \gls{iot} device can process observations locally and transmit only the data relevant to a given task, which reduces transmission energy~\cite{huang2023semantic,shiraishi2023energy}. The computation cost this adds is incurred only per inference and is offset by the transmission cost it saves. Keeping the model in \gls{dram}, by contrast, consumes energy continuously through periodic refresh, even during standby~\cite{shiraishi2025wireless}. This standby refresh cost has been overlooked in protocol design for 6G applications.

\gls{dram} approximation lengthens the refresh period to reduce this refresh energy at the cost of a controlled level of bit errors~\cite{liu2011flikker,raha2016quality,nguyen2020approximate}. Existing schemes adapt the refresh rate to data criticality~\cite{liu2011flikker} or to a user-specified quality level in smart-camera and multiview inference systems~\cite{raha2018approximating,das2024towards}. However, none of them adapts the approximation to the receiver's time-varying goals or to the relevance of the stored model to the current task, so the device keeps refreshing a model that no task requires.

To address this, the framework of wireless memory activation and approximation~\cite{shiraishi2025wireless} lets an \gls{es} remotely activate \gls{dram} only when a relevant model is needed and set its refresh period according to the required approximation level. The device then spends refresh energy only when a task requires the model, and only on that model. This framework, however, does not address the energy-efficient retrieval of timely relevant data under a deadline. In particular, it accounts for neither the semantic relevance filtering performed at the device nor the model size.  

This letter studies energy-efficient retrieval of timely and relevant data from a single \gls{iot} device by a deadline, a problem central to decision-making in applications such as rescue robotics and wildlife monitoring. We adopt the framework of~\cite{shiraishi2025wireless} and extend it as follows: \emph{i}) we design a new energy-efficient communication framework and protocol that integrate memory, computation, and communication; \emph{ii}) we introduce an analytical model of the energy consumption of the \gls{iot} device and adopt a goal-oriented metric, the \gls{goe}~\cite{agheli2024effective}, which captures the \gls{aoi}~\cite{kosta2017age} of the freshest relevant observation at the deadline; \emph{iii}) we derive the expected \gls{goe} and energy consumption in closed form, jointly select the activation timing, refresh period, model size, and threshold for semantic filtering under a freshness constraint, and quantify the energy savings over an always-on baseline. This work is the first to tailor wireless memory activation and approximation to timely relevant data retrieval and to design an energy-efficient protocol that accounts for the interaction among memory, computation, and communication. Especially, compared with~\cite{shiraishi2025wireless}, this work newly investigates optimal selection of \gls{nn} and threshold of semantic filtering for wireless memory approximation considering trade-off between energy consumption and information freshness with respect to the deadline based on our derived equations. 
\input{figure/arch.tex}
\section{System Model}\label{sec:system_model}

We consider a scenario in which an \gls{es} retrieves data from a single \gls{iot} device through a semantic query~\cite{kalor2025data}, which requires on-device inference. 
 Fig.~\ref{fig:arch} shows a configuration of the \gls{iot} device considered in this paper. 
The device has a vision sensor, a radio, non-volatile storage that holds a set of task-specific models $\mathcal{M} = \{\Phi_1, \dots, \Phi_{|\mathcal{M}|}\}$ of different sizes, and volatile \gls{dram} that holds the active model $\Phi \in \mathcal{M}$ for on-device inference with on-chip \gls{sram}.

\begin{figure}[t]
\centering
 \includegraphics[width=0.48\textwidth]{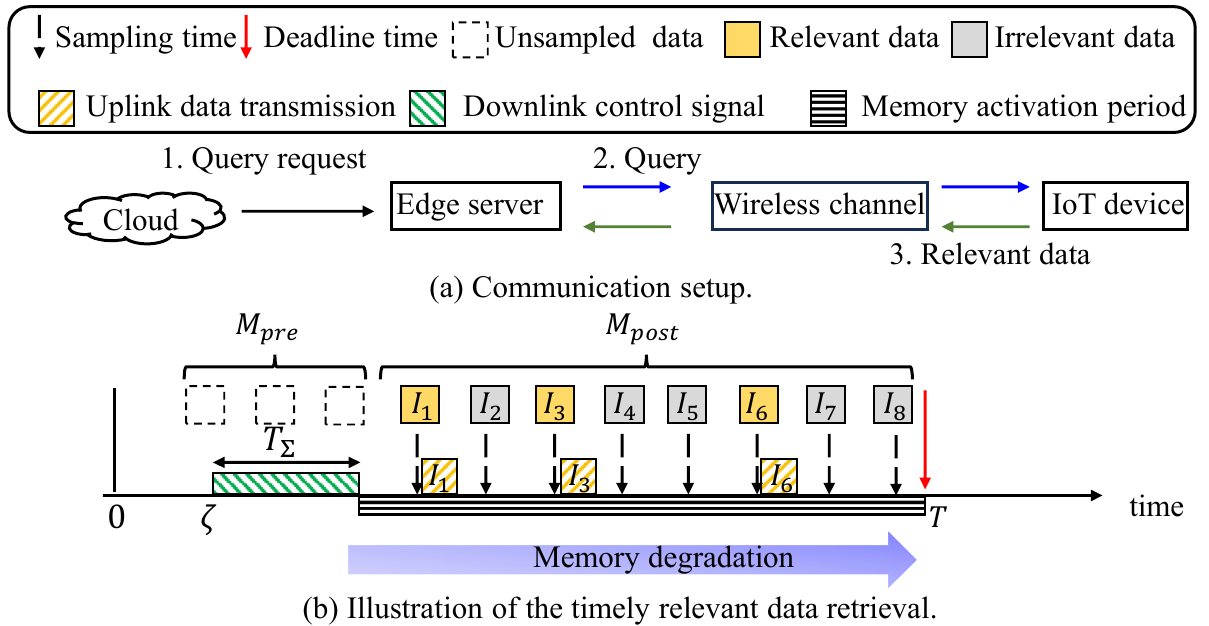}
\caption{(a) Communication setup; (b) timely relevant data retrieval from a single \gls{iot} device. Images $I_1$, $I_3$, and $I_6$ are relevant, and $I_6$ is the freshest at $T$.}\label{fig:setting}
          \vspace{-4mm}
 \end{figure}

Fig.~\ref{fig:setting}(a) shows the pull-based communication setup. Each episode is triggered by a query request from the cloud and occurs at random, infrequent instants. The request specifies the content of interest $\mathcal{X}$ (e.g., a desired bird), a deadline $T$, and a freshness requirement $\Delta_{\mathrm{th}} \in (0,\infty)$. We focus on a single episode, which starts at $t=0$. During the episode, the vision sensor samples the environment with period $T_{\mathrm{s}}$, which exceeds the time to transmit a single image.
 
To reduce the refresh energy, we adopt the wireless memory activation and approximation framework of~\cite{shiraishi2025wireless}, illustrated in Fig.~\ref{fig:setting}(b). When a new episode begins, the \gls{es} decides when to transmit the memory activation signal based on the deadline $T$, and sends it $\zeta$ seconds after the episode begins. The signal identifies the relevant model by embedding its \gls{id}. If the device detects the signal and the embedded \gls{id} matches one stored locally, it loads the corresponding model $\Phi$ from storage into \gls{dram} and begins processing its new observations, so \gls{dram} is activated on demand and consumes no refresh energy before activation. The \gls{es} then sets the refresh period to $\tau$, where a longer $\tau$ saves refresh energy at the cost of greater degradation of the model held in \gls{dram}.

After memory activation, the \gls{es} transmits the threshold $V_{\mathrm{th}}$ and a semantic query, which embeds the desired content as a feature vector $\bm{q}=\Phi_{\mathrm{ref}}(\bm{x}_{\mathrm{q}}) \in \mathbb{R}^{J}$, where $\bm{x}_{\mathrm{q}}$ is a query example and $\Phi_{\mathrm{ref}}$ is a reference model at the \gls{es}. On receiving the query, the device evaluates the relevance of each new observation with the activated model $\Phi$ and transmits the observation if its similarity measure is at least $V_{\mathrm{th}}$, as detailed in Sec.~\ref{sec:similarity_measure}. We assume the \gls{dl} and \gls{ul} channels are erasure channels, in which packet loss occurs with probability $p_{\mathrm{e}}^{\mathrm{DL}}$ and $p_{\mathrm{e}}^{\mathrm{UL}}$, respectively, without retransmission within an episode. We denote by $T_{\Sigma}$ the total overhead from the activation signal to the start of processing, which comprises the \gls{dl} reception and the model loading.   

\subsection{Observation Model}\label{sec:observation_model} 
Let $M_{\mathrm{pre}}$ and $M_{\mathrm{post}}$ be the numbers of images that the \gls{iot} device captures before and after the memory activation, respectively, given by $M_{\mathrm{pre}}=  \lfloor\frac{\zeta + T_{\Sigma}}{T_{\mathrm{s}}} \rfloor$ and $M_{\mathrm{post}} = \lfloor\frac{T-\zeta -T_{\Sigma}}{T_{\mathrm{s}}} \rfloor$. The total number of images in the episode is $M = M_{\mathrm{pre}} + M_{\mathrm{post}}$. 
The images captured after memory activation are denoted by $\mathcal{D} = [\bm{x}_{M_{\mathrm{pre}} + 1}, \ldots, \bm{x}_{M}]^\top$. Here,  ${\bm{x}_{m}} \in \mathbb{R}^{d_{\mathrm{C}}{\times}d_{\mathrm{H}}{\times}d_{\mathrm{W}}}$ is the $m$-th image, where $d_{C}$, $d_{H}$, and $d_{W}$ are the number of channels, height, and width, respectively.  

\subsubsection{\gls{dram} Retention Accuracy}\label{eq:dram_model}  
Because \gls{dram} cells store data as charge on leaking capacitors, \gls{dram} must be refreshed periodically to retain the data. Let $\tau_{\mathrm{0}}$ denote the nominal refresh period, which retains the data without error.  For $\tau > \tau_{\mathrm{0}}$, bit errors may occur in each refresh interval, as measured in~\cite{raha2016quality}. Following~\cite{shiraishi2025wireless}, we model the bit error probability of a single refresh interval as $P_e(\tau) = \tfrac{1}{2}[1 -\exp\left(-{(\tau-\tau_{\mathrm{0}})}/{\gamma}\right)] \in [0, 0.5]$, where $\gamma$ is a hardware-specific constant, and a smaller $\gamma$ gives faster memory degradation.   
The number of refresh operations between memory activation and the $m$-th sample, denoted by $w_m$, is $w_m = \sum_{m^{\prime}=M_{\mathrm{pre}}+1}^{m}{
\left\lfloor{m^{\prime}T_{\mathrm{s}}}/{\tau}\right\rfloor - \left\lfloor{(m^{\prime}-1)T_{\mathrm{s}}}/{\tau}\right\rfloor}$. 
Then, the bit error probability after $w_m$ refresh operations is $\bar{P}_{e}(w_m, \tau)  = \tfrac{1}{2}\left[1 - \left(1 -2 P_e(\tau)\right)^{w_m}\right]$.

\subsubsection{Similarity Measure}\label{sec:similarity_measure}
To infer the relevance of each observed image to the current task, the \gls{iot} device computes a similarity measure with the active model $\Phi$, as in~\cite{huang2023semantic,kalor2025data,shiraishi2023energy}. 
We assume $\Phi$ is a feature extractor that maps an image $\bm{x}$ to a feature vector $\bm{y} = \Phi(\bm{x}) \in \mathbb{R}^{J}$. Let $g(\cdot, \cdot)$ be a similarity function. Then, the observed similarity of the $m$-th image $\bm{x}_m$, denoted by $z_{m}~\in~[0, 1]$, is $z_{m} = g(\Phi(\bm{x}_m), \bm{q})$. The observed similarity $z_m$ deviates from the true similarity $\beta_m=g\left(\Phi_{\mathrm{ref}}(\bm{x}_m),\bm{q}\right)\in[0,1]$ obtained using $\Phi_{\mathrm{ref}}$. We assume that all device-side models have an embedding dimension of $J$ and their outputs are aligned with the query embedding space.

This deviation has two sources, the limited size of the device model and the \gls{dram} degradation under memory approximation~\cite{shiraishi2025wireless}, and we model the observed similarity of the $m$-th sample as
\begin{equation}
z_m = [\beta_m + \varepsilon_{m}]_0^1,~\varepsilon_{m} \sim~\mathcal{N}(0,\sigma_{\mathrm{ML}}^2(N, \tau, w_m)),\label{eq:observation_model}
\end{equation}
where $[x]_a^b = \min(\max(x,a),b)$ and the total variance $\sigma_{\mathrm{ML}}^2(N, \tau, w_m) = \sigma_{\mathrm{c}}^2(N) +  \sigma_{d,m}^2(\tau,  w_m)$ combines a static term $\sigma_{\mathrm{c}}^2(N)$ due to the limited model capacity and a time-varying term $\sigma_{d,m}^2(\tau,  w_m)$ due to memory degradation. Our experiment with the ImageNet validation set~\cite{deng2009imagenet} and MCUNet~\cite{lin2020mcunet} shows that the observed error approximately follows a Gaussian distribution whose variance increases under pruning and bit errors. We assume that $\varepsilon_m$ is independent across samples and independent of $\beta_m$. In practice, bit errors persist in the stored weights, and the errors of successive samples are positively correlated. Missed detections then cluster, and the product forms of $P_{\mathrm{G}}$ and $P_{\mathrm{F}}$ in Sec.~III give an optimistic \gls{goe}. 

We model $\sigma_{\mathrm{c}}^2$ with empirical neural scaling laws~\cite{rosenfeld2021scaling}, under which the reducible loss of a deep learning model decays as a power law in the model size and the data size. We neglect the data-dependent term, since the deployed model is trained on a large dataset, and model the capacity noise in terms of the number of parameters $N$ as 
$\sigma_{c}^2(N) = A N^{-2\alpha}$, 
where $A > 0$ is a scaling coefficient and $\alpha > 0$ is the scaling exponent. 

Next, we model $\sigma_{d,m}^2(\tau,  w_m)$. After $w_m$ refresh operations, each stored bit is in error with probability $\bar{P}_{e}(w_m, \tau)$. The bit errors perturb the quantized weights of the \gls{nn} and hence the output feature vector, and we model the resulting variance as $\sigma_{d,m}^2 = \kappa \bar{P}_{e}(w_m, \tau)(1-\bar{P}_{e}(w_m, \tau))$, 
 where $\kappa$ is an architecture-specific constant.  
 
\subsection{Grade of Effectiveness}
To quantify the freshness of the relevant data at the deadline, we adopt the \gls{goe}~\cite{agheli2024effective}, which measures the effectiveness of updates at the receiver.  
To define the \gls{goe}, we first extend the standard \gls{aoi} metric~\cite{kosta2017age}. Let $t^{\prime}_j$ and $t_j$ denote the reception and generation times of the $j$-th received image, and let $u(t)$ be the generation time of the latest image received by time $t$ whose true similarity is at least $\delta$, i.e., $u(t) = {\max}_{j: (t_j' \leq t) \land (\beta_{\lfloor{t_j}/{T_{\mathrm{s}}} \rfloor} \geq \delta)}~t_j$. 
The \gls{aoi} of relevant data at time $t$ is then $\Delta(t) = t - u(t)$, and, following the \gls{qaoi}~\cite{chiariotti2022query}, we define the \gls{goe} as $\Delta_Q= T- u(T)$. If the \gls{es} receives no relevant image by the deadline, we set $\Delta_Q = \Gamma$, as in~\cite{shiraishi_k_qaoi}. 
\subsection{Energy Model}
\subsubsection{Computation}\label{sec:comp_model} 
The \gls{nn} inference energy $E_{\mathrm{inf}}$ is the sum of the  \gls{dram} access energy ($E_{\mathrm{DRAM}}=NE_{D}$) and on-chip processing energy ($E_{\mathrm{HW}}$)~\cite{moons2017minimum}. Here, $E_{D}$ denotes the energy per $b_{q}$-bit \gls{dram} access. 

At full precision, the read/write energy $E_{L}$ of a small local \gls{sram} or register file equals the energy $E_{\mathrm{MUAC}}$ of a single \gls{muac} operation, and an access to the main \gls{sram} costs $E_{M} = 2E_{L}$~\cite{moons2017minimum}. 
Then, $E_{\mathrm{HW}} = E_{C} + E_{W} + E_{A}$, where $E_{C}=E_{\mathrm{MUAC}}(N_{c}+3A_{s})$, $E_{W}=NE_{M}+ E_{L}N_{c}/ \sqrt{p}$, and $E_{A}=2E_{M}A_{s}+E_{L}N_{c}/\sqrt{p}$ are the compute, weight-access, and activation-access energies, respectively. 
Here, $N_{c}$ is the inference complexity in terms of \gls{muac} operations and $A_{s}$ is the total number of activations across the model, and $p = 64 \times {b_{\max}}/{b_{q}}$, where $b_{\max}$ is the full-precision bit width.

\subsubsection{Communication}\label{sec:comm_cost}
Let $\mathcal{R}= \{\bm{x}_{m}| z_m \geq V_{\mathrm{th}}, \forall \bm{x}_{m}~\in~{\mathcal{D}}\}$ denote the set of images that pass the filter. With transmit and receive powers $\xi_{\mathrm{tx}}$ and $\xi_{\mathrm{rx}}$, the communication energy is $ E_{\mathrm{comm}} = \xi_{\mathrm{tx}} t_{\mathrm{UL}}|\mathcal{R}| + \xi_{\mathrm{rx}} t_{\mathrm{DL}}$. Here, $t_{\mathrm{UL}} = ({b_{q}d_{\mathrm{C}} d_{\mathrm{H}} {d_{\mathrm{W}}}})/{R_{\mathrm{UL}}}$ and $t_{\mathrm{DL}} = \left((J +3)b_{q}\right)/{R_{\mathrm{DL}}}$ are the \gls{ul} and \gls{dl} transmission times, and $R_{\mathrm{UL}}$ and $R_{\mathrm{DL}}$ are the corresponding data rates.  

\subsubsection{Memory Refresh}
The energy of a single refresh follows from its timing overhead and the \gls{dram} current (IDD) values~\cite{bhati2015flexible}.   
Let $n_{\mathrm{re}}(\zeta, \tau )$ be the number of refresh operations between memory activation and the deadline. 
Then, the total refresh energy is  $E_{\mathrm{re}}(\zeta, \tau) = \left(E_{\mathrm{AR}}+ {Nb_q}/{b_{\mathrm{row}}}E_{\mathrm{row}}\right) n_{\mathrm{re}}(\zeta, \tau )$, where $N = |\Phi|$ is the number of parameters of \gls{nn} $\Phi$, while $E_{\mathrm{AR}}$, $E_{\mathrm{row}}$ and $b_{\mathrm{row}}$ respectively represent the energy consumed by a single \gls{ar} command, that of a single row refresh, and the number of bits stored per \gls{dram} row. Loading the selected model from non-volatile storage into \gls{dram} consumes power $\xi_{\mathrm{sys}}$ over the loading time $T_{\Sigma} - t_{\mathrm{DL}}$, so the loading energy is $(T_{\Sigma} - t_{\mathrm{DL}})\xi_{\mathrm{sys}}$.  

\section{Analysis}
The \gls{iot} device transmits the $m$-th sample with probability 
\begin{equation}
P_{\mathrm{tr}}(m) =\int_{0}^{1}Q\left(\frac{ V_{\mathrm{th}}\,-\,\beta}{\sigma_{\mathrm{ML}}(N, \tau, w_m )}\right)g_{T}(\beta)d\beta,
\end{equation}
where $Q(x) = \frac{1}{\sqrt{2\pi}}\int_{x}^{\infty}\,\exp\left(-\frac{u^2}{2}\right)du$ and $g_{T}(\beta)$ is the probability density function of the true similarity. Then, the expected total energy consumption is:
\begin{equation}
\begin{split}
\mathbb{E}&[E_{\mathrm{tot}}] = (1-p_{\mathrm e}^{\mathrm{DL}}) \Bigg[ \sum_{m=M_{\mathrm{pre}}+1}^{M} P_{\mathrm{tr}}(m)t_{\mathrm{UL}}\xi_{\mathrm{tx}} +M_{\mathrm{post}}E_{\mathrm{inf}}\\ &\qquad+ \left(E_{\mathrm{AR}} +\frac{Nb_qE_{\mathrm{row}}}{b_{\mathrm{row}}}\right) \left\lfloor\frac{T-\zeta-T_\Sigma}{\tau}\right\rfloor\Bigg]+ \varrho,\label{eq:exp_ene}
\end{split}
\end{equation} 
where $\varrho = \xi_{\mathrm{rx}} t_{\mathrm{DL}}+ (1-p_{\mathrm e}^{\mathrm{DL}})(T_\Sigma-t_{\mathrm{DL}})\xi_{\mathrm{sys}}$.  

The expected \gls{goe} follows from the \gls{pmf} of the index of the freshest relevant sample that the \gls{es} receives. The $m$-th sample is relevant and delivered to the \gls{es} with probability 
\begin{equation}\begin{split}
P_{\Omega}(m) &= (1- p_{\mathrm{e}}^{\mathrm{UL}})\Pr(z_{m} \geq V_{\mathrm{th}},~\beta_{m} \geq \delta) \\&= (1- p_{\mathrm{e}}^{\mathrm{UL}}) \int_{\delta}^{1}Q\left(\frac{ V_{\mathrm{th}}\,-\,\beta}{\sigma_{\mathrm{ML}}(N, \tau, w_m )}\right)g_{T}(\beta)d\beta.
\end{split}\end{equation}
Sample $m \in [M_{\mathrm{pre}} + 1, M]$ is the freshest relevant sample at the \gls{es} if it is relevant, successfully delivered, and none of the later samples $\ell \in \{m+1, \ldots, M\}$ is both relevant and successfully delivered, so this event has probability $P_{\mathrm{G}}(m) = P_{\Omega}(m)\prod_{\ell = {m+1}}^{M} \left(1- P_{\Omega}(\ell )\right)$.
The \gls{es} receives no relevant sample, and $\Delta_Q = \Gamma$, with probability  $P_{\mathrm{F}} = \prod_{\ell =M_{\mathrm{pre}} + 1}^{M} \left(1- P_{\Omega}(\ell )\right)$.  
Finally, the expected \gls{goe} is: 
\begin{equation}\begin{split}
\mathbb{E}[\Delta_Q]&= (1- p_{\mathrm{e}}^{\mathrm{DL}})\sum_{m=M_{\mathrm{pre}}+1}^{M}(T- mT_{\mathrm{s}})  P_{\mathrm{G}}(m) \\&+ \Gamma  \left[p_{\mathrm{e}}^{\mathrm{DL}}+(1- p_{\mathrm{e}}^{\mathrm{DL}}) P_{\mathrm{F}}\right].\label{eq:exp_goe}
\end{split}\end{equation}

Using~\eqref{eq:exp_ene} and~\eqref{eq:exp_goe}, we minimize the expected total energy consumption of the \gls{iot} device subject to the \gls{goe} constraint $\Delta_{\mathrm{th}}$: 
\begin{equation}
\underset{\{\zeta, \tau, N, V_{\mathrm{th}}\}}{\min}\mathbb{E}\left[E_{\mathrm{tot}} \right]~\text{s.t.}~\mathbb{E}[\Delta_Q] ~\leq~\Delta_{\mathrm{th}}.\label{eq:min_ene_end}
\end{equation}
We solve~\eqref{eq:min_ene_end} by grid search, in which we vary $\zeta \in [ 0, 1000]$ s with a 50 s step subject to $\zeta \leq T - T_{\Sigma}$, $\tau \in [64, 1280]$ ms with a 32 ms step, $V_{\mathrm{th}} \in [0.86, 0.98]$ with a 0.04 step, and $N \in \mathcal{S} = \{100\text{K}, 250\text{K}, 500\text{K}, 1\text{M}, 2\text{M}\}$.

\section{Numerical Evaluation}
We present results for \gls{goe} and energy consumption obtained from analysis and Monte Carlo simulation over $2\times10^5$ trials, shown by lines and markers, respectively. As a baseline, we consider a semantic query with always-on memory, in which \gls{dram} is always active with the refresh period $\tau_0$ and holds the model selected for the query before the episode begins, so the baseline incurs no overhead ($T_{\Sigma} = 0$). The baseline uses the threshold $V_{\mathrm{th}}^{\star\,\mathrm{b}}$ and model size $N^{\star\,\mathrm{b}} = |\Phi^{\star\,\mathrm{b}}|$ that minimize its energy consumption under the \gls{goe} constraint. 

 Table~\ref{tab:param} lists the default parameters, which follow~\cite{khan2021empirical} for the radio, \cite{moons2017minimum} for the computation, with the energies scaled to $b_{\mathrm{q}}/b_{\max} = 0.5$, \cite{bhati2015flexible} for the \gls{dram} refresh, and \cite{lin2020mcunet} for the \gls{nn}. We take $\alpha$ from~\cite{besiroglu2024chinchilla} and set $A$ so that $\sigma_{\mathrm{c}} = 0.25$ at $N = 10^5$. For simplicity of analysis, we assume that $g_{T}(\beta)$ is uniform on $[0, 1]$ and set $\delta = 0.9$. A non-uniform $g_{T}(\beta)$ changes only the integrals in $P_{\mathrm{tr}}$ and $P_{\Omega}$. 
\begin{table}[t]
\centering
\caption{Default parameters for numerical evaluations.}
\label{tab:param}
\footnotesize
\begin{tabular}{@{}ll@{\hspace{3mm}}ll@{}}
\toprule
Parameter & Value & Parameter & Value \\
\midrule
$T$ & $10^3$~s & $T_{\mathrm{s}}$ & $10$~s \\
$\Gamma$ & $2000$ & $p_{\mathrm{e}}^{\mathrm{DL}}, p_{\mathrm{e}}^{\mathrm{UL}}$ & $0$ \\
$R_{\mathrm{UL}}, R_{\mathrm{DL}}$ & $10^6$~bps & $\xi_{\mathrm{tx}}/\xi_{\mathrm{rx}}$ & $170/160$~mW \\
$(d_{\mathrm{C}}, d_{\mathrm{H}}, d_{\mathrm{W}})$ & $(3, 128, 160)$ & $b_{\mathrm{q}}/b_{\max}$ & $8/16$ \\
$N_{c}$ & $67.3$~M & $A_{s}$ & $0.325$~M \\
$E_{\mathrm{MUAC}}$ & $1.56$~pJ & $E_{L}$ & $1.85$~pJ \\
$E_{M}$ & $3.70$~pJ & $E_{D}$ & $236.8$~pJ \\
$\tau_{0}$ & $64$~ms & $\gamma$ & $640$~ms \\
$E_{\mathrm{AR}}$ & $41.5$~nJ & $E_{\mathrm{row}}$ & $157$~nJ \\
$b_{\mathrm{row}}$ & $8$~Kbit & $\kappa$ & $0.2$ \\
$\alpha$ & $0.35$ & $A$ & $197.6$ \\
\bottomrule
\end{tabular}
\end{table}

\input{figure/zeta_Vth}
Fig.~\ref{fig_zeta_Vth_new} shows the \gls{goe} and total energy consumption of our proposed scheme as a function of memory activation time $\zeta$, where we set $V_{\mathrm{th}} = \{0.94, 0.96, 0.98\}$. The analytical and simulation results agree closely for both \gls{goe} and energy consumption, and this agreement validates the analysis. The activation time $\zeta$ trades \gls{goe} against energy consumption.
An earlier activation (small $\zeta$) gives the device more samples in which to detect and deliver a relevant image, so the \gls{goe} decreases, but the communication, computation, and refresh energy increase because all three are proportional to the active duration $T - \zeta - T_{\Sigma}$. A later activation (large $\zeta$) saves energy but increases the \gls{goe}, because the \gls{es} more often collects no relevant image. 
A higher $V_{\mathrm{th}}$ passes fewer images through the filter and lowers $P_{\Omega}$, so the \gls{goe} increases and the transmission energy decreases.

Fig.~\ref{Fig:model_size_tau_new} shows \gls{goe} and total energy consumption as a function of the number of parameters $N$, where we set $\tau = \{64, 96, 128\}$ ms. As $N$ increases, the \gls{goe} decreases and the total energy consumption increases. A larger model reduces $\sigma_{\mathrm{c}}^2$ and hence the probability of missed detection, but it occupies more \gls{dram} rows and requires more \gls{dram} accesses per inference, so the refresh and inference energy increase. A longer $\tau$ reduces the number of refresh operations, and the slope of the energy in $N$ decreases roughly as $1/\tau$. The longer $\tau$ also increases $\sigma_{d,m}^2$, which limits the benefit of a larger model: $\sigma_{\mathrm{c}}^2$ falls below $0.01$ at $N = 2$~M, so for $\tau > \tau_{0}$ the degradation term dominates $\sigma_{\mathrm{ML}}^2$ and the \gls{goe} flattens, whereas for $\tau = \tau_{0}$ the degradation term becomes zero and the \gls{goe} keeps decreasing.
\input{figure/Model_size_tau}

Fig.~\ref{fig:error_prob} shows \gls{goe} against packet error probability, in which we change the value of $p_{\mathrm{e}}^{\mathrm{DL}}$ or $p_{\mathrm{e}}^{\mathrm{UL}}$. The \gls{goe} increases with both error probabilities, but for different reasons. A \gls{dl} error prevents the device from receiving the activation signal and the query, so the episode incurs the penalty $\Gamma$, and by~\eqref{eq:exp_goe} the \gls{goe} grows linearly in $p_{\mathrm{e}}^{\mathrm{DL}}$. A \gls{ul} error loses a single image, and an earlier relevant image can still reach the \gls{es}, so the \gls{goe} grows gradually with $p_{\mathrm{e}}^{\mathrm{UL}}$. The \gls{dl} control signal therefore requires stronger protection than the \gls{ul} data. 
\begin{figure}[t]
\centering
\input{figure/impact_erasure_GoE}
\caption{\gls{goe} as a function of \gls{ul}/\gls{dl} packet error probability.}
        \label{fig:error_prob}
        \vspace{-4mm}
\end{figure}

Fig.~\ref{fig:T_sys} shows the \gls{goe} and energy consumption as a function of $T_{\Sigma}$.  A larger $T_{\Sigma}$ shortens the interval between model loading and the deadline, so the device captures fewer images after activation. The \gls{goe} therefore increases, and the inference, transmission, and refresh energy decrease. In contrast, the loading energy $(T_{\Sigma} - t_{\mathrm{DL}})\xi_{\mathrm{sys}}$ increases with $T_{\Sigma}$, and a larger $\xi_{\mathrm{sys}}$ increases the total energy.
\begin{figure}[t]
\centering
\input{figure/T_sigma_Esys_new.tex}
      \caption{\gls{goe} and energy consumption as a function of $T_{\Sigma}$.} 
        \label{fig:T_sys}
        \vspace{-4mm}
\end{figure}


To evaluate the energy efficiency of our proposed scheme, we consider the following configurations ($\gamma, \Gamma, p_{\mathrm{e}}^{\mathrm{DL}}, p_{\mathrm{e}}^{\mathrm{UL}}$): setting 1  (3200, 2000, 0, 0); setting 2 (320, 2000, 0, 0); setting 3 (3200, 5000, 0, 0); setting 4 (3200, 2000, 0, 0.1); setting 5 (3200, 2000, 0.1, 0). We obtain the optimal sets of parameters $\{\zeta^{\star}, \tau^{\star}, \Phi^{\star}, V_{\mathrm{th}}^{\star}\}$ by solving Eq.~\eqref{eq:min_ene_end}. 
Then, we define $\eta = \mathbb{E}\left[E_{\mathrm{tot}} (\zeta^{\star}, \tau^{\star}, \Phi^{\star}, V_{\mathrm{th}}^{\star})\right]/\mathbb{E}\left[E_{\mathrm{tot}} (0, \tau_{0}, \Phi^{\star\, {\mathrm{b}}}, V_{\mathrm{th}}^{\star\,\mathrm{b}})\right]$, as the ratio of the total energy consumption of the proposed scheme to that of the baseline. 

Fig.~\ref{fig:minimum_energy} shows $\eta$ against $\Delta_{\mathrm{th}}$ for the different settings. Missing points indicate that no configuration satisfies the constraint. From Fig.~\ref{fig:minimum_energy}, our proposed scheme consumes less energy than the baseline when $\Delta_{\mathrm{th}} > 150$ for settings 1 to 4. The \gls{es} achieves this saving by tuning the memory activation timing $\zeta$ and refresh period $\tau$ under the \gls{goe} constraint. Setting 2 yields a higher $\eta$ than setting 1 at every $\Delta_{\mathrm{th}}$ because its smaller $\gamma$ accelerates memory degradation and forces the \gls{es} to choose a shorter refresh period $\tau$. In setting 3, the larger penalty $\Gamma$ for failing to collect relevant data forces the \gls{es} to activate the memory earlier, so $\eta$ also exceeds that of setting 1. In setting 4, \gls{ul} losses reduce $P_{\Omega}(m)$ by the factor $1-p_{\mathrm{e}}^{\mathrm{UL}}$, so the \gls{es} must activate the memory earlier or lower $V_{\mathrm{th}}$ to deliver at least one relevant image by the deadline, and this conservative choice keeps $\eta$ above that of setting 1 at every $\Delta_{\mathrm{th}}$. The impact of channel error is larger for the \gls{dl} signaling (setting 5). In setting 5, Eq.~\eqref{eq:exp_goe} gives $\mathbb{E}[\Delta_Q] \geq \Gamma p_{\mathrm{e}}^{\mathrm{DL}} = 200$, so no configuration satisfies $\Delta_{\mathrm{th}} < 200$. For larger $\Delta_{\mathrm{th}}$, the budget left for missed detections shrinks to $\Delta_{\mathrm{th}} - 200$, and the \gls{es} must apply a smaller $V_{\mathrm{th}}$, a larger $N$, a smaller $\tau$, and an earlier $\zeta$, which increases energy consumption. 

\begin{figure}[t]
\centering
\input{figure/Minimum_Energy_new.tex}
        \caption{\BLUE{Energy consumption ratio $\eta$ against the \gls{goe} constraint $\Delta_{\mathrm{th}}$.}}
        \label{fig:minimum_energy}
        \vspace{-4mm}
\end{figure}
\section{Conclusion}
This letter studied energy-efficient, timely, and relevant data retrieval from an \gls{iot} device under wireless memory activation and approximation. We derived closed-form expressions for the expected energy consumption and the \gls{goe}. 
The numerical results characterized the energy-freshness trade-off and showed that the proposed scheme consumes less energy than the always-on baseline for relaxed freshness constraints. Future work includes an experimental feasibility study and a robust design that accounts for mismatched statistics and temporally correlated observations.

\bibliographystyle{IEEEtran}
\bibliography{IEEEabrv,Ref}

\end{document}

%% file: acronyms.tex
\newacronym{ack}{ACK}{Acknowledgement}
\newacronym{awgn}{AWGN}{Additive white Gaussian Noise}
\newacronym{ai}{AI}{Artificial Intelligence}
\newacronym{aoi}{AoI}{Age of Information}
\newacronym{ar}{AR}{Auto Refresh}
\newacronym{bs}{BS}{Base Station}

\newacronym{cts}{CTS}{Clear to Send}
\newacronym{cdf}{CDF}{Cumulative Distributed Function}
\newacronym{ccdf}{CCDF}{Complementary Cumulative Distributed Function}
\newacronym{dl}{DL}{downlink}
\newacronym{dnn}{DNN}{Deep Neural Networks}
\newacronym{dram}{DRAM}{dynamic random access memory}

\newacronym{eo}{EO}{Earth Observation}
\newacronym{es}{ES}{edge server}

\newacronym{fov}{FoV}{Field of View}
\newacronym{fp}{FP}{False Positive}
\newacronym{fn}{FN}{False Negative}

\newacronym{gs}{GS}{Ground Station}
\newacronym{goe}{GoE}{Grade of Effectiveness}
\newacronym{hmm}{HMM}{Hidden Markov Model}

\newacronym{id}{ID}{identity}
\newacronym{idwu}{IDWu}{Identity-based Wake-up}
\newacronym{iot}{IoT}{Internet of Things}

\newacronym{kpis}{KPIs}{Key Performance Indicator}
\newacronym{kkt}{KKT}{Karush-Kuhn-Tucker Conditions}
\newacronym{new_metric}{QAoI-RSEM}{QAoI for Relevant Semantic Data}

\newacronym{mcu}{MCUs}{Micro Controller Unit}
\newacronym{mac}{MAC}{Medium Access Control}
\newacronym{muac}{MUAC}{multiply-accumulation}
\newacronym{mcwu}{MCWu}{MultiCast Wake-up}
\newacronym{madd}{mADD}{multiplication and addition}
\newacronym{ml}{\textrm{ML}}{Machine Learning}
\newacronym{mlmo}{ML-Mode}{\gls{ml} Sharing Mode}
\newacronym{mlmosg}{MLMoSg}{\gls{mlmo} Signal}
\newacronym{nn}{NN}{neural network}
\newacronym{ook}{OOK}{On-Off Keying}
\newacronym{pmf}{PMF}{Probability Mass Function}
\newacronym{pdf}{PDF}{Probability Density Function}
\newacronym{qnn}{QNN}{Quantized Neural Network}

\newacronym{ppp}{PPP}{Poisson point process}
\newacronym{qaoi}{QAoI}{Query Age of Information}
\newacronym{tinyairnet}{\textrm{TinyAirNet}}{Tiny Neural Network transmission over the Air}
\newacronym{ri}{RI}{Receiver-Initiated}
\newacronym{rimosg}{RIMoSg}{RI-Mode Signal}
\newacronym{semdas}{SEMDAS}{SEMantic DAta Soucing}
\newacronym{sram}{SRAM}{static random access memory}
\newacronym{sifi}{SiFi}{Significance and Fidelity}
\newacronym{ti}{TI}{Transmitter-Initiated}
\newacronym{timosg}{TIMoSg}{TI-Mode Signal}

\newacronym{tp}{TP}{True Positive}
\newacronym{tn}{TN}{True Negative}
\newacronym{jscc}{JSCC}{Joint Source Channel Coding}

\newacronym{uav}{UAV}{Unmannded  Aerial Vehicle}
\newacronym{ul}{UL}{uplink}
\newacronym{rr}{RR}{Round-robin}

\newacronym{wsn}{WSNs}{Wireless Sensor Networks}
\newacronym{wmsn}{WMSNs}{Wireless Memory Sensor Networks}

\newacronym{wurx}{WuRx}{Wake-up receiver}

%% file: figure/arch.tex
\begin{figure}[t]
\centering
\definecolor{figamber}{RGB}{236,190,80}
\definecolor{figgray}{RGB}{200,200,200}
\definecolor{figgreen}{RGB}{110,170,110}
\definecolor{figpurple}{RGB}{120,100,180}
\begin{tikzpicture}[
scale=0.95, transform shape,
  font=\scriptsize,
  >={Latex[length=2mm]},
  blk/.style={draw, thick, rounded corners=2pt, align=center, inner sep=2pt},
  lbl/.style={font=\scriptsize, inner sep=1.5pt,
              fill opacity=0.85, text opacity=1},
  flow/.style={->, thick},
  cyl/.pic={
     \fill[figgray!55] (-0.22,-0.09) rectangle (0.22,0.09);
     \draw[thick] (-0.22,0.09)--(-0.22,-0.09); \draw[thick] (0.22,0.09)--(0.22,-0.09);
     \draw[thick, fill=figgray!55] (0,-0.09) ellipse (0.22 and 0.06);
     \draw[thick, fill=figgray!72] (0,0.09) ellipse (0.22 and 0.06);},
  cam/.pic={
     \draw[thick, fill=white] (-0.20,-0.10) rectangle (0.20,0.10);
     \draw[thick, fill=white] (0.06,0.10) rectangle (0.18,0.16);
     \draw[thick, fill=figgray!30] (0,0) circle (0.085);
     \fill[figgray!60] (0,0) circle (0.04);},
  pinsdown/.pic={\foreach \i in {0,...,6}{\draw[thick] ({-0.8+0.2667*\i},0)--({-0.8+0.2667*\i},-0.13);}},
  pinsup/.pic={\foreach \i in {0,...,6}{\draw[thick] ({-0.8+0.2667*\i},0)--({-0.8+0.2667*\i},0.13);}},
]

\draw[thick, dashed, rounded corners=5pt] (0.2,-1.08) rectangle (8.3,2.0);
\node[anchor=south west, font=\scriptsize] at (0.3,-1.03) {IoT device};

\node[blk, fill=figgray!40, minimum width=2.5cm, minimum height=0.7cm] (sto) at (1.75,1.32)
     {Storage (non-volatile)\\$\{\Phi_1,\dots,\Phi_{|\mathcal{M}|}\}$};
\node[blk, fill=figamber!70, minimum width=2.3cm, minimum height=0.7cm] (dram) at (5.6,1.32)
     {DRAM\\active NN $\Phi$};
\node[blk, fill=white, minimum width=2.3cm, minimum height=0.55cm] (sensor) at (1.75,0.14)
     {Vision sensor};
\node[blk, fill=figgreen!18, minimum width=2.3cm, minimum height=0.55cm] (cpu) at (5.6,0.14)
     {Compute + SRAM};
\node[blk, fill=white, minimum width=2.1cm, minimum height=0.4cm] (dec) at (5.6,-0.72)
     {$z_m \!\ge\! V_{\mathrm{th}}$?};

\pic at ([xshift=-11.2mm,yshift=0.8mm]sto.north)    {cyl};       
\pic at ([xshift=-10.2mm]sensor.north) {cam};       
\pic at (dram.south)                 {pinsdown};  
\pic at (cpu.north)                  {pinsup};    
\pic at (cpu.south)                  {pinsdown};
\draw[flow, black] ([xshift=0mm]dram.east) to[out=12,in=60,looseness=2]
     node[lbl, below =2mm,  xshift=8mm,text=black]{refresh ($\tau$)} ([xshift=5mm]dram.north);

\draw[flow, figgreen!70!black] (3.725,2.12) -- (3.725,1.62);
\node[lbl, anchor=south,text=figgreen!70!black,draw=figgreen!70!black,fill=white,fill opacity=1] at (3.725,2.02) {ES: control + query};
\draw[flow] (sto) -- node[lbl, above]{load $\Phi$} (dram);
\draw[flow] (sensor) -- node[lbl, above]{$\bm{x}_m$} (cpu);
\draw[flow] (dram) -- node[lbl, right]{weights} (cpu);
\draw[flow] (cpu) -- node[lbl, right]{$z_m$} (dec);
\draw[flow, black] (dec.west) -- ([xshift=-4mm]dec.west);
\node[lbl, fill=white,anchor=east,text=black,draw=black,fill opacity=1] at ([xshift=-4mm]dec.west) {transmit to ES};

\end{tikzpicture}
\caption{A configuration of the \gls{iot} device under wireless memory activation and approximation.}
\label{fig:arch}
\vspace{-4mm}
\end{figure}

%% file: figure/zeta_Vth.tex
%
\begin{figure}[t!]
\begin{subfigure}[h]{0.18\textwidth}
\centering
\begin{tikzpicture}[scale=1]

\begin{axis}[%
height=3cm,
scale only axis,
xmin=0,
xmax=800,
xlabel style={font=\color{white!15!black}},
xlabel={$\zeta$~[s]},
ymin=0,
ymax=1000,
ylabel style={font=\color{white!15!black}},
ylabel={\gls{goe}},
axis background/.style={fill=white},
legend style={at={(0,0.9)},anchor=north west,legend cell align=left, align=left, draw=none,fill=none },
legend image post style={xscale=0.8},
legend columns=1
]
\addplot[solid, color=red, line width=1pt, mark size=1pt]
  table[row sep=crcr]{%
0	187.944654885649\\
100	191.650896624464\\
200	198.539381565111\\
300	211.247184245354\\
400	234.542584265121\\
500	277.015610339356\\
600	354.090374481033\\
700	493.380860906535\\
800	744.19415623221\\
900	1194.36183356972\\
1000	2000\\
};
\addlegendentry{$V_{\mathrm{th}} = 0.94$}
\addplot [color=red, only marks, mark=square, mark options={solid, red}, forget plot]
  table[row sep=crcr]{%
0	188.4495\\
100	191.094\\
200	194.664\\
300	212.4095\\
400	237.368\\
500	277.822\\
600	355.243\\
700	490.039\\
800	749.8885\\
900	1199.2105\\
1000	2000\\
};
\addplot[dashed, color=blue, line width=1pt, mark size=1pt]
  table[row sep=crcr]{%
0	202.166256184195\\
100	206.926178859711\\
200	215.482947317802\\
300	230.750692950923\\
400	257.82091526061\\
500	305.557473811869\\
600	389.342342979287\\
700	535.792455849676\\
800	790.847053049228\\
900	1233.60954347271\\
1000	2000\\
};
\addlegendentry{$V_{\mathrm{th}} = 0.96$}
\addplot [color=blue, only marks, mark=diamond, mark options={solid, blue}, forget plot]
  table[row sep=crcr]{%
0	202.9645\\
100	205.6645\\
200	219.3585\\
300	227.4875\\
400	262.4375\\
500	313.5455\\
600	382.2775\\
700	541.7445\\
800	797.784\\
900	1245.007\\
1000	2000\\
};
\addplot[dotted, color=brown, line width=1pt, mark size=1pt]
  table[row sep=crcr]{%
0	218.654701175604\\
100	224.731228272585\\
200	235.29888756972\\
300	253.540192381532\\
400	284.828608919027\\
500	338.205430049194\\
600	428.836325811693\\
700	582.089275536965\\
800	840.291646718174\\
900	1273.90774569275\\
1000	2000\\
};
\addlegendentry{$V_{\mathrm{th}} = 0.98$}
\addplot [color=brown, only marks, mark=pentagon, mark options={solid, brown}, forget plot]
  table[row sep=crcr]{%
0	217.6655\\
100	221.8325\\
200	239.6895\\
300	256.5695\\
400	281.416\\
500	335.642\\
600	425.258\\
700	582.4085\\
800	842.2015\\
900	1267.874\\
1000	2000\\
};
\end{axis}
\end{tikzpicture}%
\label{subFig:zeta_Vth_goe}
\end{subfigure}
\hspace{2.2em}
\begin{subfigure}[h]{0.18\textwidth}
\centering
\begin{tikzpicture}[scale=1]

\begin{axis}[%
height=3cm,
scale only axis,
xmin=0,
xmax=800,
xlabel style={font=\color{white!15!black}},
xlabel={$\zeta$~[s]},
ymin=0,
ymax=1.5,
ylabel style={font=\color{white!15!black}},
ylabel={$\mathbb{E}\left[E_{\mathrm{tot}} \right]$ [J]},
axis background/.style={fill=white},
legend style={at={(0,0.42)},anchor=north west,legend cell align=left, align=left, draw=none,fill=none },
legend image post style={xscale=0.8},
legend columns=1
]
\addplot[solid, color=red, line width=1pt, mark size=1pt]
  table[row sep=crcr]{%
0	1.3713971223886\\
100	1.23439366489974\\
200	1.09737446591088\\
300	0.960371008422024\\
400	0.823351809433164\\
500	0.686348351944303\\
600	0.549329152955442\\
700	0.412325695466582\\
800	0.275306496477721\\
900	0.138303038988861\\
1000	0.00128384\\
};
\addlegendentry{$V_{\mathrm{th}} = 0.94$}
\addplot [color=red, only marks, mark=square, mark options={solid, red}, forget plot]
  table[row sep=crcr]{%
0	1.3704803498429\\
100	1.23275345741661\\
200	1.09905505005032\\
300	0.964428174264027\\
400	0.821555054577738\\
500	0.686096772711448\\
600	0.550384607905158\\
700	0.411245578518869\\
800	0.274434620752579\\
900	0.13800288352629\\
1000	0.00128384\\
};
\addplot[dashed, color=blue, line width=1pt, mark size=1pt]
  table[row sep=crcr]{%
0	1.27460871948395\\
100	1.14728410228556\\
200	1.01994374358716\\
300	0.892619126388767\\
400	0.765278767690372\\
500	0.637954150491977\\
600	0.510613791793581\\
700	0.383289174595186\\
800	0.25594881589679\\
900	0.128624198698395\\
1000	0.00128384\\
};
\addlegendentry{$V_{\mathrm{th}} = 0.96$}
\addplot [color=blue, only marks, mark=diamond, mark options={solid, blue}, forget plot]
  table[row sep=crcr]{%
0	1.2740456004029\\
100	1.14781216589661\\
200	1.01931109101032\\
300	0.892488569784027\\
400	0.765107177457738\\
500	0.635539762791448\\
600	0.510543962785158\\
700	0.384732498198869\\
800	0.254439095632579\\
900	0.12667654240629\\
1000	0.00128384\\
};
\addplot[dotted, color=brown, line width=1pt, mark size=1pt]
  table[row sep=crcr]{%
0	1.18307996810587\\
100	1.06490822604529\\
200	0.946720742484698\\
300	0.82854900042411\\
400	0.710361516863523\\
500	0.592189774802936\\
600	0.474002291242349\\
700	0.355830549181762\\
800	0.237643065621175\\
900	0.119471323560587\\
1000	0.00128384\\
};
\addlegendentry{$V_{\mathrm{th}} = 0.98$}
\addplot [color=brown, only marks, mark=pentagon, mark options={solid, brown}, forget plot]
  table[row sep=crcr]{%
0	1.1819600656829\\
100	1.06296278861661\\
200	0.947250326850317\\
300	0.827864503224027\\
400	0.709649467377737\\
500	0.591257988711448\\
600	0.475269784225158\\
700	0.357041244438869\\
800	0.237217709392579\\
900	0.12053917792629\\
1000	0.00128384\\
};
\end{axis}
\end{tikzpicture}%
\label{subFig:zeta_Vth_energy}
\end{subfigure}

\caption{\BLUE{\gls{goe} and energy consumption as a function of $\zeta$.}}
\vspace{-4mm}
\label{fig_zeta_Vth_new}
\end{figure}

%% file: figure/model_size_tau.tex
%
%
\begin{figure}[t!]
\begin{subfigure}[t]{0.2\textwidth}
\centering
\begin{tikzpicture}[scale=1]

\begin{axis}[%
height=3cm,
scale only axis,
xmin=10^5,
xmax=2*10^6,
xtick={500*10^3,10^6,1.5*10^6,2*10^6},
 xticklabels={
 \(\displaystyle 0.5\),
  \(\displaystyle 1\),
  \(\displaystyle 1.5\),
  \(\displaystyle 2\),
  },
scaled x ticks = false,
xlabel style={font=\color{white!15!black}},
xlabel={$N$ [M]},
ymin=100,
ymax=200,
ylabel style={font=\color{white!15!black}},
ylabel={\gls{goe}},
axis background/.style={fill=white},
legend style={at={(0.2,1)},anchor=north west,legend cell align=left, align=left, draw=none,fill=none },
 legend image post style={xscale=0.8},
legend columns=1
]
\addplot[solid, color=red, line width=1pt, mark size=1pt]
  table[row sep=crcr]{%
100000	164.945141421153\\
250000	156.283694394493\\
500000	148.777112938748\\
1000000	140.653795795878\\
2000000	132.251820982934\\
};
\addlegendentry{$\tau = 64$ ms}
\addplot [color=red, only marks, mark=square, mark options={solid, red}, forget plot]
  table[row sep=crcr]{%
100000	163.9625\\
250000	157.931\\
500000	148.1115\\
1000000	139.0635\\
2000000	131.6625\\
};
\addplot[dashed, color=blue, line width=1pt, mark size=1pt]
  table[row sep=crcr]{%
100000	166.671733450208\\
250000	159.963646483447\\
500000	154.925759920091\\
1000000	150.30983031059\\
2000000	146.403289851987\\
};
\addlegendentry{$\tau = 96$ ms}
\addplot [color=blue, only marks, mark=diamond, mark options={solid, blue}, forget plot]
table[row sep=crcr]{%
100000	166.871\\
250000	160.818\\
500000	152.2375\\
1000000	151.559\\
2000000	146.1285\\
};
\addplot[dotted, color=brown, line width=1pt, mark size=1pt]
  table[row sep=crcr]{%
100000	167.327544470479\\
250000	161.234503956042\\
500000	156.851839128687\\
1000000	153.00185455142\\
2000000	149.87709552701\\
};
\addlegendentry{$\tau = 128$ ms}
\addplot [color=brown, only marks, mark=pentagon, mark options={solid, brown}, forget plot]
  table[row sep=crcr]{%
100000	167.528\\
250000	160.841\\
500000	155.16\\
1000000	154.789\\
2000000	147.7825\\
};

\end{axis}
\end{tikzpicture}%
\label{subFig:model_size_tau_goe}
\end{subfigure}
\hspace{1.2em}
\begin{subfigure}[t]{0.2\textwidth}
\centering
\begin{tikzpicture}[scale=1]

\begin{axis}[%
height=3cm,
scale only axis,
xtick={500*10^3,10^6,1.5*10^6,2*10^6},
 xticklabels={
 \(\displaystyle 0.5\),
  \(\displaystyle 1\),
  \(\displaystyle 1.5\),
  \(\displaystyle 2\),
  },
scaled x ticks = false,
xlabel style={font=\color{white!15!black}},
xlabel={$N$ [M]},
ymin=0,
ymax=7,
ylabel style={font=\color{white!15!black}},
ylabel={$\mathbb{E}\left[E_{\mathrm{tot}} \right]$ [J]},
axis background/.style={fill=white},
legend style={at={(0.25,0.4)},anchor=north west,legend cell align=left, align=left, draw=none,fill=none },
 legend image post style={xscale=0.8},
legend columns=1
]
\addplot[solid, color=red, line width=1pt, mark size=1pt]
  table[row sep=crcr]{%
100000	1.58038274590806\\
250000	1.74716490753516\\
500000	2.25899209451136\\
1000000	3.42270146087561\\
2000000	5.85127957859392\\
};
\addlegendentry{$\tau = 64$ ms}
\addplot [color=red, only marks, mark=square, mark options={solid, red}, forget plot]
  table[row sep=crcr]{%
100000	1.5820335069629\\
250000	1.7454478838429\\
500000	2.2536626890429\\
1000000	3.4256822181629\\
2000000	5.8481241231229\\
};
\addplot[dashed, color=blue, line width=1pt, mark size=1pt]
  table[row sep=crcr]{%
100000	1.53511561962883\\
250000	1.58905343227345\\
500000	1.90338881031259\\
1000000	2.66346917653888\\
2000000	4.27652518275767\\
};
\addlegendentry{$\tau = 96$ ms}
\addplot [color=blue, only marks, mark=diamond, mark options={solid, blue}, forget plot]
  table[row sep=crcr]{%
100000	1.5355692430629\\
250000	1.5897812840229\\
500000	1.9041923573029\\
1000000	2.6624882446629\\
2000000	4.2801021249829\\
};
\addplot[dotted, color=brown, line width=1pt, mark size=1pt]
  table[row sep=crcr]{%
100000	1.51008915805381\\
250000	1.50663830340105\\
500000	1.72146663715483\\
1000000	2.27929277956225\\
2000000	3.48495083762867\\
};
\addlegendentry{$\tau = 128$ ms}
\addplot [color=brown, only marks, mark=pentagon, mark options={solid, brown}, forget plot]
  table[row sep=crcr]{%
100000	1.5097400487429\\
250000	1.5092958500229\\
500000	1.7206810509829\\
1000000	2.2789664660229\\
2000000	3.4859516628229\\
};
\end{axis}
\end{tikzpicture}%
%
\label{subFig:model_size_tau_energy}
\end{subfigure}

\caption{\BLUE{\gls{goe} and energy consumption as a function of model size.}} 
 \vspace{-4mm}
\label{Fig:model_size_tau_new}
\end{figure}

%% file: figure/impact_erasure_GoE.tex
\begin{tikzpicture}

\begin{axis}[%
width=7 cm,
height=3cm,
scale only axis,
xmin=0,
xmax=0.25,
xlabel style={font=\color{white!15!black}},
xtick={0, 0.05, 0.1, 0.15, 0.2, 0.25},
 xticklabels={
 \(\displaystyle 0\),
  \(\displaystyle 0.05\),
  \(\displaystyle 0.10\),
  \(\displaystyle 0.15\),
  \(\displaystyle 0.20\),
  \(\displaystyle 0.25\),
  },
scaled x ticks = false,
xlabel={$p_{\mathrm{e}}$},
ymin=150,
ymax=650,
ylabel style={font=\color{white!15!black}},
ylabel={GoE},
axis background/.style={fill=white},
legend style={at={(0,1)},anchor=north west,legend cell align=left, align=left, draw=none,fill=none },
 legend image post style={xscale=0.8},
]
\addplot [color=red,line width=1pt]
  table[row sep=crcr]{%
0	164.945141421153\\
0.05	174.781385402037\\
0.1	185.87455329563\\
0.15	198.480346463594\\
0.2	212.920799874653\\
0.25	229.60356727428\\
};
\addlegendentry{$(p_{\mathrm{e}}^{\mathrm{DL}}, p_{\mathrm{e}}^{\mathrm{UL}})= (0, p_{\mathrm{e}})$}

\addplot [dotted,color=blue,line width=1pt]
  table[row sep=crcr]{%
0	164.945141421153\\
0.05	256.697884350096\\
0.1	348.450627279038\\
0.15	440.20337020798\\
0.2	531.956113136923\\
0.25	623.708856065865\\
};
\addlegendentry{$(p_{\mathrm{e}}^{\mathrm{DL}}, p_{\mathrm{e}}^{\mathrm{UL}})= (p_{\mathrm{e}}, 0)$}

\addplot [color=red, only marks, mark=o, mark options={solid, red}]
  table[row sep=crcr]{%
0	162.844\\
0.05	176.6555\\
0.1	185.483\\
0.15	198.0015\\
0.2	213.306\\
0.25	228.7895\\
};

\addplot [ color=blue, only marks, mark=+, mark options={solid, blue}]
  table[row sep=crcr]{%
0	164.604\\
0.05	260.696\\
0.1	354.188\\
0.15	435.142\\
0.2	533.902\\
0.25	620.5785\\
};

\end{axis}

\end{tikzpicture}%

%% file: figure/T_sigma_Esys_new.tex
\begin{tikzpicture}

\begin{axis}[
    name=leftaxis,
    width=5cm,
    height=3cm,
    xmode=log,
    log basis x=10,
    xmin=10,
    xmax=1000,
    xtick={10,20,50,100,200,500,1000},
    xticklabels={10,20,50,100,200,500,1000},
    xlabel={$T_{\Sigma}$},
    ylabel={GoE},
    tick align=outside,
    axis y line*=left,
    axis x line*=bottom,
    legend style={
        at={(0.4,0.2)},
        anchor=south,
        legend columns=1,
        draw=none,fill=none,
    },
    legend image post style={xscale=0.8}
]
\draw[black] (rel axis cs:0,1) -- (rel axis cs:1,1);
\addplot[
dashdotted,
    color=blue,
    line width=1pt
]
table{
10      165.104946117563
20      165.276246925318
50      165.867503766984
100     167.170431281361
200     171.590747842102
500     230.653819422384
1000    2000
};
\addlegendentry{GoE}

\addplot[
    color=blue,
    only marks,
    mark=square,
    mark options={solid, blue},
    forget plot
]
table{
10      163.6355
20      163.351
50      166.8645
100     167.452
200     170.766
500     235.283
1000    2000
};

\addplot[
    color=blue,
    only marks,
    mark=diamond,
    mark options={solid, blue},
    forget plot
]
table{
10      164.1205
20      165.9395
50      164.941
100     165.746
200     171.677
500     228.167
1000    2000
};

\addplot[
    color=blue,
    only marks,
    mark=pentagon,
    mark options={solid, blue},
    forget plot
]
table{
10      165.903
20      165.471
50      165.731
100     165.852
200     174.623
500     230.7475
1000    2000
};

\addlegendimage{solid,red,line width=1pt}
\addlegendentry{Energy $(\xi_{\mathrm{sys}}=1\,\mu\mathrm{W})$}

\addlegendimage{dashed,red,line width=1pt}
\addlegendentry{Energy $(\xi_{\mathrm{sys}}=100\,\mu\mathrm{W})$}

\addlegendimage{dotted,red,line width=1pt}
\addlegendentry{Energy $(\xi_{\mathrm{sys}}=1\,\mathrm{mW})$}

\end{axis}

\begin{axis}[
    at={(leftaxis.south west)},
    anchor=south west,
    width=5cm,
    height=3cm,
    xmode=log,
    log basis x=10,
    xmin=10,
    xmax=1000,
    xtick={10,20,50,100,200,500,1000},
    xticklabels=\empty,
    xlabel={},
    ylabel={$\mathbb{E}[E_{\mathrm{tot}}]$ [J]},
    axis y line*=right,
    axis x line=none,
    tick align=outside,
]

\addplot[
    solid,
    color=red,
    line width=1pt
]
table{
10      1.56460568419998
20      1.5488286305159
50      1.50148172796366
100     1.42258071804326
200     1.26476295670245
500     0.791341155680032
1000    0.002283831976
};

\addplot[
    color=red,
    only marks,
    mark=square,
    mark options={solid, red}
]
table{
10      1.56171921930827
20      1.55111442575764
50      1.49873311512575
100     1.42442361515261
200     1.26438733002632
500     0.792268895247448
1000    0.002283831976
};

\addplot[
    dashed,
    color=red,
    line width=1pt
]
table{
10      1.56559488982398
20      1.5508078361399
50      1.50643093358766
100     1.43247992366726
200     1.28456216232645
500     0.840840361304032
1000    0.1012830376
};

\addplot[
    color=red,
    only marks,
    mark=diamond,
    mark options={solid, red}
]
table{
10      1.56326408829227
20      1.55197394882164
50      1.50781846154975
100     1.43183278045661
200     1.28341779837032
500     0.841450578951448
1000    0.1012830376
};

\addplot[
    dotted,
    color=red,
    line width=1pt
]
table{
10      1.57458766822398
20      1.5688006145399
50      1.55142371198766
100     1.52247270206726
200     1.46455494072645
500     1.29083313970403
1000    1.001275816
};

\addplot[
    color=red,
    only marks,
    mark=pentagon,
    mark options={solid, red}
]
table{
10      1.57188085389227
20      1.56722601170164
50      1.54727549594975
100     1.52109442285661
200     1.46438821005032
500     1.28978472311145
1000    1.001275816
};

\end{axis}

\end{tikzpicture}

%% file: figure/Minimum_Energy_new.tex
\definecolor{mycolor1}{rgb}{0.00000,0.44700,0.74100}%
\definecolor{mycolor2}{rgb}{0.85000,0.32500,0.09800}%
\definecolor{mycolor3}{rgb}{0.92900,0.69400,0.12500}%
\definecolor{mycolor4}{rgb}{0.49400,0.18400,0.55600}%
\begin{tikzpicture}

\begin{axis}[%
width=7cm,
height=3cm,
scale only axis,
unbounded coords=jump,
xmin=100,
xmax=500,
xlabel style={font=\color{white!15!black}},
xlabel={$\Delta_{\mathrm{th}}$},
ymin=0.1,
ymax=4,
ylabel style={font=\color{white!15!black}},
ylabel={$\eta$},
axis background/.style={fill=white},
legend style={at={(0,1)},anchor=north west,legend cell align=left, align=left, draw=none,fill=none },
 legend image post style={xscale=0.8},
legend columns=3
]
\addplot [solid, color=red, line width=1pt, mark size=1pt]
  table[row sep=crcr]{%
100	nan\\
150	0.473811378502311\\
200	0.41720413145353\\
250	0.416384528060934\\
300	0.341420678151352\\
350	0.311273385885878\\
400	0.270018987580377\\
450	0.241021847241326\\
500	0.229394116779911\\
};
\addlegendentry{set. 1}

\addplot [dashed, color=blue, line width=1pt, mark size=1pt]
  table[row sep=crcr]{%
100	nan\\
150	0.717570130280926\\
200	0.596518780602143\\
250	0.546184877656096\\
300	0.467876238627118\\
350	0.40515645559721\\
400	0.376443419983806\\
450	0.331467722727096\\
500	0.285507756453234\\
};
\addlegendentry{set. 2}

\addplot [dotted, color=brown, line width=1pt, mark size=1pt]
  table[row sep=crcr]{%
100	nan\\
150	0.550492321747868\\
200	0.522047478182478\\
250	0.588903364185092\\
300	0.539106453714977\\
350	0.481931378988823\\
400	0.438477520660808\\
450	0.421623248420691\\
500	0.390912665297102\\
};
\addlegendentry{set. 3}

\addplot [dashdotted,color=magenta, line width=1.0pt]
  table[row sep=crcr]{%
100	nan\\
150	0.723357993175277\\
200	0.440855087238513\\
250	0.483612740261364\\
300	0.409773623111751\\
350	0.360086666072365\\
400	0.316379137208523\\
450	0.290084108536281\\
500	0.26216108436214\\
};
\addlegendentry{set. 4}

\addplot [color=cyan, line width=1.0pt]
  table[row sep=crcr]{%
100	nan\\
150	nan\\
200	nan\\
250	nan\\
300	3.75101198714092\\
350	0.530697440167617\\
400	0.400164991196368\\
450	0.350395418744545\\
500	0.291126197719889\\
};
\addlegendentry{set. 5}

\end{axis}
\end{tikzpicture}%